\documentclass[12pt]{article}
\usepackage{tikz}
\usetikzlibrary{positioning, arrows}
\usepackage{cancel}
\usepackage{amssymb}
\usepackage{amsfonts}
\usepackage{amsmath}
\usepackage{xcolor}
\usepackage{caption}

\numberwithin{equation}{section}

\usepackage{graphicx}

\usepackage[affil-it]{authblk}
\newcommand{\mb}[1]{{\bss{#1}}}

\newcommand{\bs}[1]{{\boldsymbol{#1}}}

\newcommand{\dsl}[1]{{\displaystyle{#1}}}

\newcommand{\eps}{\epsilon}

\usepackage{graphicx}
\newcommand{\HH}{{{\mathcal{H}}}}
\newtheorem{theorem}{Theorem}[section]

\newtheorem{remark}[theorem]{Remark}
\newenvironment{rem}{\begin{remark} \rm}{\end{remark}}

\newcommand{\wit}[1]{{{\widetilde{#1}}}}
\newcommand{\ou}{{\overline{u}}}

\newcommand{\BB}{\mathcal{B}}

\newcommand{\bm}[1]{\mbox{\boldmath{$#1$}}}
\newcommand{\bss}[1]{\boldsymbol{#1}}

\renewcommand{\sfdefault}{bch}
\def\barr{\hbox{{\fontfamily{\sfdefault}\selectfont I\hskip -.35ex R}}}
\def\sbarr{\hbox{{\fontfamily{\sfdefault}\selectfont {\scriptsize I}\hskip -.25ex {\scriptsize R}}}}

\newcommand{\RR}{{\sbarr}}

\newcommand{\Asf}{{\mathsf A}}
\newcommand{\Bsf}{{\mathsf B}}
\newcommand{\Csf}{{\mathsf C}}

\newcommand{\Sigb}{{{\bs{\Sigma}}}}

\begin{document}
\title{ A Hamiltonian {\color{black}  approach to} the CH-KP equation}
\author{The authors}
\author{ 
R. Camassa${}^1$, G. Falqui${}^{2,3}$,  E. Sforza${}^{2,3,4}$
\medskip\\
{\small $^1$University of North Carolina, Carolina Center for Interdisciplinary Applied Mathematics,}
\\ 
{\small Department of Mathematics, Chapel Hill, NC 27599, USA}
\medskip\\
{\small  $^2$Department of Mathematics and Applications, University of  Milano-Bicocca,}
\\
{\small Via Roberto Cozzi 55, I-20125 Milano, Italy} 
\medskip\\
{\small  $^3$INFN, Sezione di Milano-Bicocca, Piazza della Scienza 3, I-20126 Milano, Italy}\\ \medskip
{\small  ${}^4$Joint Ph.D. program University of Milano - Bicocca, University of Pavia and INdAM}\\ \medskip
}

\medskip

\maketitle
\abstract{\noindent  
A model governing quasi-unidirectional wave propagation at the surface of a shallow layer of water is derived from Euler equations by long wave asymptotics together with Hamiltonian reduction techniques. The balance between nonlinearity, dispersion and weak dependence on the second planar coordinate is examined and results in the so-called CH-KP equation, a mildly nonlinear version of the well known Kadomtsev-Petviashvili equation. The Hamiltonian structure of the model is obtained by Dirac reduction from the intermediate step of a fully-nonlinear, long-wave system for stratified fluids. A particular scaling is considered that reduces such system to the one-dimensional CH equation in a limiting case. Examples of weak peakon-type solutions are provided, chosen to illustrate the collision behavior supported by the limiting equation.
}


\section{Introduction}
The Kadomtsev-Petviashvili (KP) equation is a nonlinear scalar partial differential equation in two spatial  and one temporal coordinate which describes the evolution of nonlinear, long water waves  of small amplitude with negligible surface tension and weak  dependence on the transverse  spatial coordinate  $y$; this equation with appropriate scaling and reference frame choice can be normalized to 
{\color{black} 
\begin{equation}\label{KPII}
(u_t + 6 u u_x + u_{xxx})_x + u_{yy }= 0\, .
\end{equation}
It is regarded  as being a {\em universal} integrable equation, since a great number of integrable effective models in one space dimension can be obtained by  suitable reductions of the KP equation.  The most obvious one is the celebrated Korteweg-de Vries (KdV) equation, directly obtained by  imposing $y$-independence of the dependent variable $u$ in~\eqref{KPII} {\color{black} (see, e.g., \cite{Dikii} and the references quoted therein).}
}

Among shallow water integrable PDEs in one spatial dimension, the so-called CH equation~\cite{CH,CHH}  plays a distinguished role, in that, with respect to the KdV and Boussinesq-type  equations, allows for a higher order mixing of nonlinearity and dispersion, and presents a number of interesting features such as the presence of elastically interacting $N$-peaked solutions, wave breaking phenomena, and others. It also admits a biHamiltonian formulation, and can be shown to  correspond to the geodesic  motion of a right-invariant Sobolev metric on the Virasoro group,  a universal central extension of 
the diffeomorphism group on the unit circle $ \mathrm{Diff} (S^{1})$\cite{AK2021,KM03}.
Two dimensional extensions of the CH equations have a comparatively short history. To the best of our knowledge, a multi-parameter family of extensions of the CH equation  first appeared in~\cite{Chen06} in context of nonlinear elasticity theory. The derivation of the 2D extension of the CH equation within water wave theory was obtained in \cite{Guietal}, (and later further studied  in \cite{GLPe24}), by means of a suitable expansion of the classical 3D water wave equation entailing, specifically,  a weak dependence on the transversal coordinate $y$.
 Such equation was called CH-KP equation, as its extension in the transverse $y$ direction parallels that of the KP extension of KdV.
 
This paper aims first at providing an alternative derivation of the CH-KP equation, based on the Hamiltonian picture of water wave theory. In particular, we use the geometric setting of~\cite{CFOPS25} where a  reduction of the Hamiltonian structure of~\cite{Ben86,  Bow87} for heterogeneous incompressible fluid flows 
 was used to cast the dynamics of 3D two-layer sharply stratified fluids into a Hamiltonian  formalism. We remark that the natural variables arising from the process are the interface graph and the horizontal momentum evaluated at the interface.
The 2D {\color{black} model, variously referred to as Serre, Su \& Gardner or Green \& Naghdi \cite{Ser56, SG69, GN76} (herafter SGN)} equations, can be readily obtained in such a formalism, simply by letting the density of the upper fluid go to zero while extending its unperturbed width go to infinity.  
Once the notion of weak transversality is introduced, and a CH-like expansion in a small amplitude parameter is performed, we obtain a set of weakly transversal Hamiltonian SGN equations. Changing the momentum coordinate from the interface value to a value $u^*$ at a specific height (see~\cite{Johnson,CL09} for a similar approach), and by a unidirectionalization process, yields an evolution equation of CH-type, the Hamiltonian structure for this being obtained by means of a Dirac reduction process. Further manipulations with asymptotic equivalences yield the actual CH-KP equation governing the evolution of the dependent  CH-like variable $m=u^*-u^*_{xx}$ and the associated Hamiltonian operators.

The second main aim of this paper is to study and present families of  solutions of the CH-KP equation, especially in the case where the vanishing of  a smoothing parameter yields peakons as solitary wave solutions of the CH-KP equations. Mimicking the CH ansatz~\cite{CH,CHH} for $N$-peakons solution, we generalize to the CH-KP case the finite-dimensional Hamiltonian system governing their evolution.  Finally, the case $N=2$ is studied in some details, and different $2$-peakon  collisions are illustrated.

The layout of the paper is as follows: in Section~\ref{sect:SGN} we briefly review the results of \cite{CFOPS25} about the Hamiltonian structure for  $2$-layer fluids, perform the air-water limit and recover  the SGN equations in this formalism. In Section~\ref{WTCHKP} we discuss the weakly transversal asymptotics for the SGN equations and discuss how the  unidirectionalization of such a system leads to the CH-KP equation. Section~\ref{Peaksol} is devoted to the discussion of a few case-studies of peakon solutions and their scattering.
We collect in Appendix~A a brief sketch of the theory of Dirac brackets and in Appendix~B details on the computations
needed to obtain the undirectionalization of  Section~\ref{WTCHKP}  and provide its  Hamiltonian structure. 

\section{The  $\mathbf{2D}$ SGN equations and their Poisson-Darboux  representation}\label{sect:SGN}
This section is devoted to describe how a special canonical representation of the SGN equations can be obtained starting from a suitable Hamiltonian structure of the Euler equations for stratified fluids.

The water wave system (with no surface tension) is usually written in the Euler form 
\begin{equation}\label{wweq}
\left\{
\begin{array}{ll}
\Delta \phi=0&\text {in } \barr^2\times (-h,\zeta)\times[0,T)\vspace{0.2cm}\\
\zeta_t=\phi_z-\zeta_x\phi_x-\zeta_y\phi_y& z=\zeta \vspace{0.2cm}\\
\phi_t=-g\zeta-\frac12|\bm{\nabla}\phi|^2&\ z=\zeta \vspace{0.2cm}\\
\phi_y=0&\ z=-h\, ,
\end{array}
\right.
\end{equation}
for the velocity potential $\phi(x,y,z;t)$ and the wave profile $\zeta(x,y;t)$, in an horizontally unbounded domain, the vertical variable $z$ being bounded by the flat bottom $z=-h$. In the seminal paper \cite{Zak68} it was discovered that \eqref{wweq} admit a canonical Hamiltonian structure, with canonical coordinates given by the wave profile $\zeta$ and the ``trace" of the potential $\phi$, that is the variable
\begin{equation}
\label{psidef}
\psi(x,y;t):=\phi(x,y, \zeta(x,y;t);t)\, .
\end{equation}

In~\cite{CFOPS25} we presented Hamiltonian representation for sharply-stratified $2$-layer configurations in $3$ dimensions, based on the so-called Bowman-Benjamin's (BB) Hamiltonian representation~\cite{Ben86, Bow87} for general incompressible flows with variable density. Here we shall use such a representation - to be briefly summarized below - specializing ot to the case in which the density of the upper fluid vanishes, while its asymptotic heights becomes unbounded.
Such a picture deduces (a descendent of) Zakharov's Hamiltonian structure by means of a Poisson reduction of the Hamiltonian structure (BB structure) devised Benjamin~\cite{Ben86} and Bowman~\cite{Bow87} for heterogenous incompressible flows.
The basic features of such a picture are the following:
\begin{enumerate}
\item The Hamiltonian variables are the density $\rho$ and the ``weighted vorticity vector"
\begin{equation}\label{Sigmadef}
\bs{\Sigma}=\bm{\nabla}\times (\rho\,\bs{U})\, .
\end{equation}
\item The Hamiltonian  operator is  the operator-valued $4\times4$ matrix in the evolution equations 
\begin{equation}
\label{PBow}
\left(
\begin{array}{c}\bigskip
\rho_t\\
\Sigb_t
\end{array}
\right)=
-\left(
\begin{array}{cc}\bigskip
0&\bm{\nabla}\cdot(\rho\, \bm{\nabla}\times\>\> )\\
\bm{\nabla}\times(\rho\, \bm{\nabla}\cdot \>\>)
&\bm{\nabla}\times(\mb{\Sigma}\times \bm{\nabla}\times \> \>)
\end{array}\right)\, \left(
\begin{array}{c}\medskip
\dsl{\frac{\delta \HH}{\delta\rho}}\\
\dsl{\frac{\delta \HH}{\delta \Sigb}}
\end{array}
\right)\, , 
\end{equation}
which can be identified with the Lie-algebraic Poisson tensor associated with semi-direct product 
$
\mathcal{X}^0(\RR^3)\ltimes C^\infty(\RR^3)
$
of the space of divergenceless vector fields times the space of functions in $\RR^3$.
\item The Hamiltonian functional is  the total energy
\begin{equation}
\label{Hbow}
H=\int_\BB \left(\frac12\rho |\mb{U}|^2+g\, z(\rho-\rho_0) \right)d^3{x}\, , 
\end{equation}
\end{enumerate}
This structure can be specialized to the case of a sharply stratified two-layered configuration, in which a homogeneous fluid of constant density $\rho_2$ lies under another fluid of constant density $\rho_1$ (with 
$\rho_2>\rho_1$ for stability). The two fluids are separated  by an  interface  $z=\zeta(x,y,t)$, i.e., $\zeta$ is a function of the horizontal variables $(x,y)$, and evolves with time $t$. The Euler variables are expressed via the Heaviside $\theta$-function as
\begin{equation}\label{no1}
\begin{split}
&\rho(x,z,t)=\rho_2+(\rho_1-\rho_2)\theta(z-\zeta(x,y,t))\\
&\mb{U}(x,y,z,t)=\mb{U}_2(x,y,z,t)+(\mb{U}_1(x,y,z,t)-\mb{U}_2(x,y,z,t))\, \theta(z-\zeta(x,y,t)),\\
\end{split}
\end{equation}
where $\mb{U}_1=(u_1,v_1,w_1)$ and $ \mb{U}_2=(u_2,v_2,w_2)$ denote the velocity vector fields in the two domains, while the reference far-field density
is
\begin{equation}\
\label{rho03D}
\rho_0(z)=\rho_2+(\rho_1-\rho_2)\theta(z),\quad z\in(-h_2,h_1)\, .
\end{equation}
Assuming the flow to be potential in the bulk of both domains, i.e., 
\begin{equation}\label{pot}
\mb{U}_i=\bm{\nabla}{\phi}_i\, ,\quad i=1,2,
\end{equation}
the characteristic variable { $\Sigb$} turns out to be a ``vortex sheet'' supported on the interface  
given by
\begin{equation}
\label{Sigmasheet}
\Sigb(x,y,z;t)=\delta(z-\zeta)\, 
\left(
\begin{array}{c}
\rho_2 v_2-\rho_1v_1+\zeta_y(\rho_2 w_2-\rho_1w_1)\\
-\left(\rho_2 u_2-\rho_1u_1+\zeta_x(\rho_2 w_2-\rho_1w_1)\right)\\
\zeta_x(\rho_2 v_2-\rho_1v_1)-\zeta_y(\rho_2 u_2-\rho_1u_1)
\end{array}
\right)
=\delta(z-\zeta)\, \left(
\begin{array}{c}
	\mu_2(x,y,t)\\  \mu_1(x,y,t)\\ {\chi}(x,y,t)
\end{array}
\right),
\end{equation}
where the seemingly odd notation relies on the fact that $\mu_i$ represent indeed the horizontal components of the momentum shear.
It was proven in~\cite{CFOPS25} that it holds:
\begin{enumerate}
\item The third component $\chi$ of the vortex sheet can be expressed as the cross product
\begin{equation}\label{chieq}
\chi= \zeta_y \mu_1 - \zeta_x \mu_2\, , 
\end{equation}
where  $(\mu_1,\mu_2)$ are the cartesian components of the $2$-vector $\bm{\mu}$. Hence the triplet of functions \begin{equation}\label{zetamucoords}
(\zeta, \mu_1,\mu_2)
\end{equation}
of the independent horizontal variables $(x,y)$
provides a natural set set of free coordinates for the 2-layered configurations.
\item   $\bs{\mu}$ is curl-free, and abusing notation a little,
\begin{equation}\label{0curv}
\nabla\times\mb{\mu}\equiv  \mu_{1\,y}-\mu_{2\,x}=0\, .
\end{equation}
\item The BB Hamiltonian structure~\eqref{PBow} can be Poisson-reduced  to the space of 2-layered configurations to the simple tensor compactly expressed in the coordinates~\eqref{zetamucoords} via 
\begin{equation}
   P= \begin{pmatrix}\label{poisson_tensor_v}
        0&-\nabla \cdot\\
        -\nabla & 0
    \end{pmatrix}\, ,
\end{equation}
where henceforth $\nabla$ refers to the 2-dimensional operator.
\end{enumerate}
These results do not depend either on the densities $\rho_j$  or on the asymptotic heights $h_j$, and hence hold true in the limit 
\begin{equation}
\rho_1\to0, \quad h_1\to\infty\, , 
\end{equation}
that is, in the physical setting of the water wave system, where, setting $\rho_2=1$ and dropping all subscripts referring to the lower (second) layer, (i.e., henceforth $h_2=h, u_2=u$, 
etc.) 
we have 
\begin{equation}\label{darvww}
\begin{cases}
\mu_1&=\wit{u}+\zeta_x \wit{w}\\
\mu_2&=\wit{v}+\zeta_y \wit{w}\, ,
\end{cases}
\end{equation}
and the Hamiltonian~\eqref{Hbow} reads
\begin{equation}\label{HWW}
H=\int_{\RR^2} dx\, dy\int_{-h}^{\zeta}\left(\frac{1}2  (u^2+v^2+w^2)+g(\eta-h) z \right)\, dz\, .
\end{equation}
\subsection{The energy for the long wave SGN model}
To obtain effective long wave models, we adimensionalize the independent variables as
\begin{equation}\label{scalingxyz}
      x = Lx^*, \quad y =Ly^*, \quad z = {h}z^* \, , 
  \end{equation}
where  $L$ denotes the typical horizontal wavelength of the dynamics, 
 and the unperturbed vertical  water thickness $h=$ offers a natural vertical scale. 
Long wave asymptotics is defined assuming that the parameter $\epsilon ={{h}}/{L}$ is small. 
As it is well known (see, e.g.,~\cite{Wh2000,Wu00}), the potential ${\phi}$ can be expressed as the Taylor series
\begin{equation}\label{phiexp}
     {\phi}(x,y,z;t) = \sum_{n=0}^{\infty}  \frac{(-1)^n}{(2n)!}(1+z)^{2n}\eps^{2n}(\partial_x^{2}+\partial_y^{2})^n\varphi(x,y;t)\,  .
\end{equation}
where we dropped asterisks and denoted by $\varphi(x,y;t)$ the potential at the bottom $z=-h$.
The expansion induces a corresponding long wave asymptotic expansions for the velocity $\bs{U}=(\bs{u},w)$ which read, at $O(\eps^2)$ 
\begin{equation}\label{veloc}
\begin{split}
\bs{u}&=\bs{u}_0-\dfrac{\eps^2}{2} (1+z)^2\Delta\bs{u}_0+O(\eps^4)\\
w&=-\eps( 1+z)\nabla\cdot\bs{u}+O(\eps^3)\, ,
\end{split}
\end{equation}
where $\bs{u}_0=\nabla\varphi(x,y)$ is the horizontal velocity at the bottom.  The surface velocities are thus expressed as
\begin{equation}
\begin{split}\label{suveloc}
\bs{\wit{u}}&=\bs{u}_0-\dfrac{\eps^2}{2} \eta^2\Delta\bs{u}_0+O(\eps^4)\\
\wit{w}&=-\eps \eta\nabla\cdot\bs{u}_0+O(\eps^3)\, ,
\end{split}
\end{equation}
Inverting the near to identity operator $\mb{1}-\dfrac{\eps^2}2\eta^2\Delta$ as $\mb{1}+\dfrac{\eps^2}2\eta^2\Delta$ we express
the bottom horizontal velocities and the surface vertical velocity in terms of the horizontal surface ones as
\begin{equation}
\begin{split}\label{su-bvel}
\bs{u}_0&=\bs{\wit{u}}+\dfrac{\eps^2}{2} \eta^2\Delta\bs{\wit{u}}+O(\eps^4)\\
\wit{w}&=-\eps \eta\nabla\cdot\bs{\wit{u}}+O(\eps^3)\, .
\end{split}
\end{equation}
Performing the nested integral  in ~\eqref{HWW} at $O(\eps^2)$ we get
\begin{equation}\label{T0}
H=\frac{1}2\int_{\RR^2}\left(\eta |\bs{u}_0|^2-\frac{\eps^2}{3}\eta^3(\bs{u}_0\cdot\Delta\bs{u}_0+(\nabla\cdot\bs{u}_0)^2+g (\eta-1)^2\right)\, dx \,dy\, , 
\end{equation}
which reads, using~\eqref{su-bvel}
\begin{equation}\label{Ttilde}
H=\frac{1}2\int_{\RR^2}\left(\eta |\bs{\wit{u}}|^2+\frac{\eps^2}{3}\eta^3(2 \bs{\wit{u}} \cdot\Delta\bs{\wit{u}}+(\nabla\cdot\bs{\wit{u}})^2+g (\eta-1)^2\right)\, dx\, dy\, , 
\end{equation}
a result matching that of~\cite{Mat16}.

The next task is to write the Hamiltonian  in terms of the triple $(\eta, \bs{\mu})$. To this end we notice that, thanks to the second of~\eqref{suveloc}, in the long-wave $O(\eps^2)$ asymptotics we have from~\eqref{darvww}
\begin{equation}\label{utomu}
\bs{\mu}=\bs{\wit{u}}-\eps^2\eta (\nabla\cdot\bs{\wit{u}})\, \nabla\eta+O(\eps^4)\, , 
\end{equation}
that can be asymptotically inverted as
\begin{equation}\label{mutou}
\bs{\wit{u}}=\bs{\mu}+\eps^2\eta (\nabla\cdot\bs{\mu})\, \nabla\eta+O(\eps^4)\, .
\end{equation}
Inserting this in~\eqref{Ttilde} we obtain
\begin{equation}\label{Ttilde2}
H=\frac{1}2\int_{\RR^2}\left(\eta |\bs{\mu}|^2+2\eps^2 (\nabla\cdot\bs{\mu})\eta^2 \nabla\eta\cdot \bs{\mu}) +\frac{2\eps^2}{3}\eta^3( \bs{\mu} \cdot\Delta\bs{\mu})+\frac{\eps^2}{3}\eta^3(\nabla\cdot\bs{\mu})^2+g (\eta-1)^2\right)\, dx\, dy\, .
\end{equation}
Observing that
\begin{equation}\label{vectrel}
3(\nabla\cdot\bs{\mu})\eta^2 \nabla\eta\cdot \bs{\mu}+\eta^3( \bs{\mu} \cdot\Delta\bs{\mu})+\eta^3(\nabla\cdot\bs{\mu})^2=
\nabla\left(\eta^3(\nabla\cdot\bs{\mu})\bs\mu\right) -\eta^3 \bs{\mu} \cdot \nabla\times(\nabla\times\bs{\mu})\, 
\end{equation}
and taking into account the irrotationality condition~\eqref{0curv}, we obtain the final simple effective form of the Hamiltonian as
\begin{equation}\label{H2DSGN}
H=\frac{1}2\int_{\RR^2}\left(\eta |\bs{\mu}|^2-\frac{\eps^2}{3}\eta^3(\nabla\cdot\bs{\mu})^2+g (\eta-1)^2\right)\, dx\, dy\, .
\end{equation}
Using the expression~\eqref{poisson_tensor_v} for the Poisson operator, 
the ensuing Hamiltonian equations are
\begin{equation}\label{Hsgneq}
\left\{
\begin{array}{lcl}\medskip
\eta_t=-\nabla\cdot \dfrac{\delta H}{\delta \bs{{\mu}}}\\
\bs{{\mu}}_t=-\nabla \Big(\dfrac{\delta H}{\delta \eta}\Big)\, , 
\end{array}\right.
\end{equation}
where
\begin{equation}\label{gradH}
\dfrac{\delta H}{\delta \bs{{\mu}}}=\eta\bs{{\mu}}+\dfrac{\eps^2}{3}\nabla(\eta^3 (\nabla\cdot \bs{{\mu}})), \quad 
\dfrac{\delta H}{\delta \eta}=\frac12\left(|\bs{{\mu}}|^2-\eps^2\eta^2(\nabla\cdot\bs{{\mu}})^2+g(\eta-1)\right)\, .
\end{equation}
We remark that  the irrotationality condition $\nabla\times\bs{{\mu}}=0$ is preserved (provided it is satisfied by the initial conditions), thanks to the functional form of the $\bs{\mu}$ evolution equations~\eqref{Hsgneq}.

%
%
%
%
{\color{black} System \eqref{Hsgneq} is  the canonical version of the famous SGN system. To recover its standard formulation (see, e.g., \cite{Mat16})  we consider, along with $\eta$, the vector of averaged horizontal velocities~(see, e.g.,~\cite{Wu81,Wu98})
\begin{equation}
\bs{\bar u}=\frac{1}{\eta}
\int_{-1}^{\zeta} \bs {u}(x,y,z)dz\,.
\end{equation} 
By using equations \eqref{veloc},  \eqref{su-bvel} and \eqref{utomu}, 
we get the relation
\begin{equation}
\bs{\bar u} = \bs{\mu}+\frac{\eps^2}{3 \eta} \nabla (\eta^3 (\nabla \cdot \bs{\mu})). 
\end{equation}
This expression can be asymptotically inverted as 
\begin{equation}\label{ubartomu}
\bs{\mu}=\bs{\bar{u}}-\frac{\eps^2}{3\eta}\nabla(\eta^3 (\nabla\cdot\bs{\bar{u}}))+O(\eps^4)\, .
\end{equation}
{\color{black} Hence, system~\eqref{Hsgneq}, which explicitly reads
\begin{equation}
\left\{
\begin{array}{lcl}\medskip
\eta_t=-\nabla\cdot \left( \eta\bs{{\mu}}+\dfrac{\eps^2}{3}\nabla(\eta^3 (\nabla\cdot \bs{{\mu}})\right)
\\
\bs{{\mu}}_t=-\nabla\left(\frac12\left(|\bs{{\mu}}|^2-\eps^2\eta^2(\nabla\cdot\bs{{\mu}})^2+g\eta \right)\right) \, , 
\end{array}\right.
\end{equation}
%
 acquires, recalling that $\nabla\times\bs{\mu}=0$, the usual form of the SGN system }\begin{equation}
\label{SGNeq}
\left\{
\begin{array}{l}\medskip
\dsl{\eta_t+\nabla\cdot(\eta\bs{\bar{u}})}=0\,  \\
\dsl{\bs{\bar{u}}_t+(\bs{\bar{u}}\cdot\nabla)\bs{\bar{u}}
+g \nabla \eta-\frac{\eps^2}{3\, \eta}\nabla\left(\eta^3(\nabla \cdot \bs{\bar{u}}_{t}+(\bs{\bar{u}}\cdot\nabla)(\nabla \cdot \bs{\bar{u}})-(\nabla \cdot \bs{\bar{u}})^2  \right)=0}\, .
\end{array}
\right.\,  
\end{equation}
This coordinate change has a clearcut Hamiltonian interpretation. The energy \eqref{H2DSGN} is written as
\begin{equation}
\label{H2DSGN1}
H=\frac{1}2\int_{\RR^2}\left(\eta |\bs{\bar u}|^2+\frac{\eps^2}{3}\eta^3(\nabla\cdot\bs{\bar u})^2+g (\eta-1)^2\right)\, dx\, dy\, .
\end{equation}
The standard Hamiltonian structure for the system \eqref{SGNeq} (see e.g.,~\cite{CHLev,Mat16}), make use of the momentum variable $\bs{m}$ defined as
\begin{equation}
\bs{m}=\dfrac{\delta H}{\delta \bs{\bar{u}}}\,,
\end{equation}
explicitly
\begin{equation}
\bs{m}= \eta \bs{\bar{u}}-\frac{\eps^2}{3}\nabla(\eta^3 (\nabla\cdot\bs{\bar{u}}))+O(\eps^4)\,.
\end{equation}
Since 
\begin{equation}
\bs{m} = \eta \bs{\mu} + O(\eps^4)\,, 
\end{equation} 
the canonical Hamiltonian structure \eqref{poisson_tensor_v} becomes, in the variables $(\eta, \mb{m})$, the SGN one
\begin{equation}\label{poissontensor}
\begin{split}
H &=  \frac{1}{2} \int \left(\bs{m} \cdot \bar{\bs{u}} + g(\eta-1)^2 \right)dx\, dy\\
{P}&=-\begin{pmatrix}
\medskip
0&\partial_x\eta&\partial_y \eta\\ \medskip
\eta\partial_x&m_1\partial_x+\partial_x m_1&\partial_y m_1 + m_2\partial_x \\
\eta \partial_y& m_1 \partial_y+\partial_x m_2 &m_2\partial_y+\partial_y m_2
\end{pmatrix}\, .
\end{split}
\end{equation}

}

\section{The weakly transversal 2D SGN equations and the CH-KP model}\label{WTCHKP}
The weakly transversal $2D$ SGN model is obtained from the  $2D$ SGN model in a standard way, by considering a relative scaling in the  $y$ variable {\color{black} \begin{equation}
y=\beta y'
\end{equation} 
with
\begin{equation}
\beta\ll \eps
\end{equation} 
 so that $y$-derivatives are of order $\beta=o(\eps)$.} Since $\bs{\mu}$ is potential, the new hamiltonian variable 
$\bs{\mu}$ is related with the one used in Section~\ref{sect:SGN} one by
\begin{equation}
\mu_1=\mu'_1,\quad \mu_2=\beta\mu_2'\, .
\end{equation}
Substituting in the Hamiltonian~\eqref{H2DSGN} and dropping primes we obtain the weakly transversal SGN Hamiltonian
\begin{equation}\label{weaktrans}
{H}_{w}= \frac{1}{2}  \int_{\RR^2}\Big( \eta (\mu_1^2 +\beta^2 \mu_2^2)-\frac{\eps^2}{3} \eta^3 \mu_{1\, x}^2 + g(\eta-1)^2\Big) dx dy \, ,
\end{equation}
engendering the motion equations
\begin{equation}\label{wtsgneq}
\left\{
\begin{array}{l}\smallskip
\eta_t+ (\eta\mu_1)_x+\dfrac{\eps^2}{3}(\eta^3\mu_{1\, x})_{xx}+\beta^2(\eta\mu_2)_y=0\\ \smallskip
\mu_{1 t}+\dfrac12(\mu_1^2+\beta^2\mu_2^2)_x-\dfrac{\eps^2}{2}(\eta^2\mu_{1\,x}^2)_x+g\eta_x =0\\ \smallskip
\mu_{2 t}+\dfrac12(\mu_1^2+\beta^2\mu_2^2)_y-\dfrac{\eps^2}{2}(\eta^2\mu_{1\,x}^2)_y+g\eta_y =0
\end{array}
\right.
\end{equation}
according with the Poisson tensor~\eqref{poisson_tensor_v}, explicitly written as
\begin{equation}\label{pxy}
P=-\begin{pmatrix}
0&\partial_x&\partial_y\\
\partial_x&0&0\\
\partial_y&0&0
\end{pmatrix}
\end{equation}
To obtain the CH-KP equations in this setting  we shall introduce, along with the two small dispersion parameters $\eps$ and $\beta$, the amplitude small parameter $\alpha=a/h$, where $a$ denotes a typical wave amplitude, and define
\begin{equation}\label{etazeta}
\eta=1+\alpha\zeta\, .
\end{equation}
We assume the scaling relations
\begin{equation}
 \eps^2 \ll\alpha { \color{black}\ll \eps\,,  \quad \beta \ll \alpha},
\end{equation}
and asymptotically expand the Hamiltonian~\eqref{weaktrans} retaining terms of order $\alpha\eps^2$ and discarding terms of $\alpha^2\eps^2, \alpha\beta^2, \eps^4$ and higher, obtaining
\begin{equation}\label{Hwtas}
H_{w}^a[\zeta,\mu_1,\mu_2]= \frac{1}{2}  \int_{\RR^2} \Big((1+\alpha\zeta)\, \mu_1^2 +\beta^2 \mu_2^2-\frac{\eps^2}{3} (1+{\color{black}3}\alpha\zeta) \mu_{1\, x}^2 + g\zeta^2\Big) dx dy \, .
\end{equation}
We thus obtain the new  asymptotic system
\begin{equation}\label{wtsgneqalpha}
\left\{
\begin{array}{lcl}\smallskip
\zeta_t+ \big((1+\alpha\zeta)\mu_1\big)_x+\dfrac{\eps^2}{3}\big((1+{\color{black}3}\alpha\zeta)\mu_{1\, x}\big)_{xx}+\beta^2\mu_{2\,y}=0\\ \smallskip
\mu_{1 t}+{\color{black}\alpha}\Big(\mu_1\mu_{1\,x}-\dfrac{{\color{black}\alpha}\eps^2}{2}{\color{black}}\mu_{1\,x}^2\Big)_x+ g \zeta_x =0\vspace{0.2cm}\\ \smallskip
\mu_2-\partial_x^{-1}\mu_{1\, y}=0 \, , 
\end{array}
\right.
\end{equation}
where we consistently traded the third equation in ~\eqref{wtsgneq} for the irrotationality constraint.

Following [citiamo chi?] we now pass to the natural Hamiltonian variables to a new set, where we basically trade the  $x$-component of the surface momentum $\mu_1$ for a velocity computed at a specific $\zeta$-dependent height.  

To this end, we notice that, by using equations~\eqref{suveloc}, \eqref{su-bvel}, \eqref{mutou}, we get the general relation
\begin{equation}\label{mutovel}
\bs{u}(x,y,z)=\bs{\mu}+\dfrac{\eps^2}{2}(1-(1+z)^2)\Delta\bs{\mu}+\alpha\eps^2\, \nabla(\zeta(\nabla\cdot\bs{\mu}))\, , 
\end{equation}
whence, for its first component we get, still in the $\beta=o(\eps)$ asymptotics, 
\begin{equation}
\label{mu1tou}
u(x,y,z)=\mu_1+\dfrac{\eps^2}{2}(1-(1+z)^2){\mu}_{1\, xx}+{\alpha\eps^2}\, (\zeta\mu_{1\, x})_x\, .
\end{equation}
We now define the new horizontal longitudinal velocity variable via
\begin{equation}\label{ust3d}
u^*(x,y):= u\big(x,y,  (1+\alpha \zeta(x))/\sqrt{{2}}-1)\big)\, , 
\end{equation}
By substituting in \eqref{mu1tou}, we  find the  asymptotic relation between $u^*$ and $\mu_1$
\begin{equation}\label{mu2ustar}
u^*= \mu_1 + \frac{\epsilon^2}{4}\mu_{1\, xx}+\alpha \epsilon^2\left(\zeta_x\mu_{1\, x}+\frac{1}{2}\zeta \mu_{1\, xx}\right)\, , 
\end{equation} 
with asymptotic inverse
\begin{equation}\label{ust2mu}
\mu_1=u^*-\frac{\epsilon^2}{4}u^*_{ xx}-\alpha \epsilon^2\left(\zeta_x u^*_{x}+\frac{1}{2}\zeta u^*_{xx}\right)\, .
\end{equation}
Since the Jacobian of the coordinate transformation $(\zeta, \mu_1,\mu_2)\to (\zeta, u^*, \mu_2)$ is
\begin{equation}
\label{Jaco*}
J=\begin{pmatrix}
1&0&0\\
\alpha\eps^2\widehat{A}&1+\dfrac{\eps^2}{4}\partial_x^2+\alpha\eps^2\widehat{B}&0\\
0&0&1
\end{pmatrix}
\end{equation}
with
\begin{equation}
\label{ABops}
\widehat{A}=\dfrac12 (\partial_x\mu_{1,x}+\mu_{1,x}\partial_x), \quad \widehat{B}=\zeta_x\partial_x+\dfrac12\zeta\partial_x^2\, , 
\end{equation}
the Poisson tensor in the new coordinates is given by the expansion
\begin{equation}\label{npoit}
\wit{P}=P-\eps^2\begin{pmatrix}
0&\partial_x^3&0\\\partial_x^3&0&0\\0&0&0
\end{pmatrix}
-\alpha\eps^2\begin{pmatrix}
0&\partial_x\widehat{B}^T&0\\
{\color{black}\widehat{B}\partial_x}&\partial_x\wit{A}+\wit{A}\partial_x&\wit{A}\partial_y\\
0&\partial_y\wit{A}^T&0
\end{pmatrix}\, , 
\end{equation}
where $\wit{A}=\dfrac12 (\partial_xu^*_x+u^*_{x}\partial_x)$. The Hamiltonian~\eqref{Hwtas} transforms into
\begin{equation}
\label{nHtwas}
H_w^a[\zeta,u^*]=\frac{1}{2}  \int_{\RR^2} \left((1+\alpha\zeta) {u^*}^2 +\beta^2 (\partial_x^{-1}u^*_y)^2
+\dfrac{\eps^2}6{u_x^*}^2+\dfrac{\alpha \epsilon^2}{2} (\zeta{u^*_x}^2-\zeta_x u^*u^*_x) 
 + g\zeta^2\right)\,  dx dy
\end{equation}
where we used the constraint $\mu_{1 y}=\mu_{2_x}$ and~\eqref{ust2mu} to obtain the non-local term, and the ensuing equations of motion for the variables $(\zeta, u^*)$ are 
\begin{equation}
\label{Hequzeta}
\left\{
\begin{array}{l}\medskip
\zeta_t+u^*_x+\alpha (\zeta u^*)_x+\dfrac{\epsilon^2}{12}u^*_{xxx}+{\color{black}\alpha \dfrac{\epsilon^2}{4}(\zeta_x u^*_{xx}+\zeta u^*_{xx})}+\beta^2\partial_x^{-1}u^*_{yy}=0\\


u^*_t+\zeta_x+\alpha u^*u^*_x+\dfrac{\epsilon^2}{4}\zeta_{xxx}+{\color{black}\alpha \dfrac{\epsilon^2}{2}(2 u^*_xu^*_{xx}+2 \zeta_{x}\zeta_{xx}+\zeta \zeta_{xxx})}=0
\, .
\end{array}
\right.
\end{equation}

%

{\color{black}Unidirectionalization can be obtained ~\cite{CL09,Guietal} imposing the 
(asymptotic) relation
 \begin{equation}\label{2dcostr2}
 \zeta -( u^* +\frac{\alpha}{4}{u^*}^2 -\frac{\epsilon^2}{12}u^*_{xx}-\alpha \epsilon^2 (\frac{17}{48}{u^*_{x}}^2+\frac{5}{24}u^*u^*_{xx}))=0\, .
 \end{equation}
To frame such a procedure within the Hamiltonian formalism, we cosider the new set of coordinates 
 $( u^*,\Phi)$, where $\Phi$ is defined by the LHS of \eqref{2dcostr2}.
 
 A long but straightforward computation shows that, in this further coordinate system, the Poisson operator reads
\begin{equation}\label{Pnew0}
P^{\mathrm{c}}=\begin{pmatrix} P^c_{11}& P^{\mathrm{c}}_{12}\\ -{P^{\mathrm{c}}_{12}}^T&P^{\mathrm{c}}_{22}\end{pmatrix}
\end{equation}
where
\begin{equation}\label{Pnew}
\begin{split}
P^{c}_{11}&=
{\color{black}\alpha \epsilon^2 
 \frac{1}{2}(\partial_x u^*_{xx} + u^*_{xx} \partial_x)}
\, .
\\
P^{c}_{12}&=-\left(\partial_x + \frac{\epsilon^2}{4} \partial_x^3
+ \alpha \epsilon^2 {\color{black}\Big(-
 \partial_x^2 u^*_x  +\frac{1}{2} \partial_x^3 u^*
  -\frac{1}{2}(u^*_{xx}\partial_x + \partial_x u^*_{xx})
\Big)}\right)\, ,
\\
P^c_{22}&=2 \partial_x + \frac{\alpha}{2}(u^* \partial_x + \partial_x u^*) + \frac{\epsilon^2}{3} \partial_x^3
+  \alpha \epsilon^2{\color{black}
   + \frac{5}{12}(\partial_x^3 u^* + u^* \partial_x^3)
}\, ,
\end{split}
\end{equation}
Performing the steps of the Dirac  reduction procedure, we use formula \eqref{pdr} to compute 
\begin{equation}
P_{red} = P^c_{11}-P^c_{12}(P^c_{22})^{-1}P^c_{21}\,,
\end{equation}
 where
\begin{equation}
(P^c_{22})^{-1} = \frac{\partial_x^{-1}}{2}-\frac{\alpha}{8}(\partial_x^{-1}u^*+u^*\partial_x^{-1})-\frac{\epsilon^2}{12}\partial_x+\alpha \epsilon^2 \left(\frac{1}{48}(u^*\partial_x +\partial_x u^*)-\frac{1}{12}(\partial_x^2 u^*\partial_x^{-1}+\partial_x^{-1}u^*\partial^2)\right)\, ,
\end{equation}
is the asymptotic inverse operator of $P^c_{22}$ (see Appendix~A for more details).
The final result for the Poisson tensor on the submanifold defined by the constraint $\Phi=0$ and thus parametrized by $u^*$ is
\begin{equation}\label{redtensor}
P_{red} = -\frac{\partial_x}{2} +\frac{\alpha}{8}(\partial_x u^* +u^* \partial_x)-\frac{\epsilon^2}{6} \partial_{x}^3 +\alpha \epsilon^2 \left(-\frac{1}{8}(\partial_x^3 u^*+u^*\partial_x^3)+\frac{47}{96}(\partial_{x}u^*_{xx}+u^*_{xx}\partial_x)\right)\,.
\end{equation}
The Hamiltonian \eqref{nHtwas} restricted to the constraint \eqref{2dcostr2} is 
\begin{equation}\label{Hrestr}
\mathcal{H}= \int \Big({u^*}^2 + \alpha \frac{3}{4}{u^*}^3 +\frac{\epsilon^2}{6} {u^*_x}^2 +\alpha \epsilon^2 \frac{3}{16}{u^*_x}^2 u^*+\frac12 \beta^2 (\partial_x^{-1}u_y)^2\Big) dxdy\, .
\end{equation}
The reduced Poisson structure gives the evolution equations 
\begin{equation}\label{CL11}
u^*_t +u^*_x +\frac{3}{2} \alpha u^*u^*_x+\frac{\epsilon^2}{6}u^*_{xxx}+\alpha \epsilon^2\left(\frac{35}{24} u^*_x u^*_{xx}+\frac{5}{12}u^* u^*_{xxx}\right)+\frac{\beta^2}{2}\partial_x^{-1}u_{yy}=0
\end{equation}
which can be rewritten in a local way as
\begin{equation}\label{CL1/2}
\left(u^*_t +u^*_x +\frac{3}{2} \alpha u^*u^*_x+\frac{\epsilon^2}{6}u^*_{xxx}+\alpha \epsilon^2(\frac{35}{24} u^*_x u^*_{xx}+\frac{5}{12}u^* u^*_{xxx})\right)_x+\frac{\beta^2}{2}u_{yy}=0\,.
\end{equation}
To obtain the CH-KP equation one can trade derivatives thanks to the asymptotic equivalence (see, e.g.,~\cite{ChCa99})   
\begin{equation}
u^*_{xxt} = -u^*_{xxx} -\frac{3}{2}\alpha (u^*u^*_x)_{xx} +O(\eps^2)\, , 
\end{equation}
so that the third derivative term in \eqref{CL11}  can be split into
\begin{equation}\label{CL2}
\frac{\epsilon^2}{6} u^*_{xxx}\equiv \frac{5}{12} \epsilon^2 u^*_{xxx} -\frac{\epsilon^2}{4}u^*_{xxx} = -\frac{5}{12} \epsilon^2 u^*_{xxt}-\frac{5}{8}(3 u^*_x u^*_{xx}+u^* u^*_{xx}) -\frac{\epsilon^2}{4}u^*_{xxx}\,. 
\end{equation}
The Hamiltonian evolution equation \eqref{CL1} is therefore asymptotically equivalent to 
\begin{equation}\label{CL31}
\left(u^*_t -\frac{5}{12}\epsilon^2 u^*_{xxt}+u^*_x +\frac{3}{2} \alpha u^*u^*_x-\frac{\epsilon^2}{4}u^*_{xxx}-\alpha \epsilon^2 \frac{5}{24}(2 u^*_x u^*_{xx}+u^* u^*_{xxx})\right)_x +\frac{\beta^2}{2} u_{yy} =0\,,
\end{equation}
which coincides with the one obtained in~\cite{Guietal} by a direct inspection of \eqref{GNuzeta}.
To arrive at a canonical form of the CH-KP equation one  performs the scaling
\begin{equation}\label{resstartocan}
u^*(x,t)= a U(b(x-vt),dy,ct)
\end{equation}
with
\begin{equation}\label{abc}
a =\frac{2(1-v)}{{\color{black}2\hat{k}\alpha}},\quad b^2 = \frac{12}{5{\color{black}\,\eps^2}}, \quad v = \frac{3}{5}, \quad c =\frac{b(1-v)}{2\hat{k}}, \quad d = \frac{\sqrt{2 b(1-v)}}{{\color{black}\beta}}\, ,
\end{equation}
with $ \hat{k}>0$.\\
Indeed, a tedious but straightforward computation shows that the new variable $U$ satisfies 
\begin{equation}\label{CH_KP}
\left(U_t- U_{xxt} +2\hat{k}U_x+3  UU_x \right)_x =\left( UU_{xxx}+2U_x U_{xx}\right)_x - 2 \hat{k} U_{yy}\, .
\end{equation}
The equation \eqref{CH_KP} has a nonlocal form 
\begin{equation}
\label{CH-KP}
U_t- U_{xxt} +2\hat{k}U_x+3  UU_x  = ( UU_{xxx}+2U_x U_{xx}) -2 \hat{k} \partial_x^{-1}U_{yy}\, ,
\end{equation}
which also possesses a canonical bihamiltonian structure analogously to the CH equation, its  1D counterpart. Indeed, as also noted in \cite{Guietal}, the following quantities
\begin{equation}
\mathcal{E} = \frac12 \int \Big(U^2 + U_x^2\Big) dxdy \,,\quad \mathcal{F} = \frac{1}{2}\int \hat{k} \Big((U^2+(\partial_x^{-1}U_{y})^2)+ U^3 + UU_x^2 \Big)dx dy\,,
\end{equation}
are conserved by \eqref{CH_KP}. If we define, as in \cite{CHH}, 
\begin{equation}
m = U-2U_{xx}\,,
\end{equation}
the equation \eqref{CH_KP} has two  Hamiltonian structures,
\begin{equation}
m_t + P_1 \frac{\delta\mathcal{ E}}{ \delta m} =0\, \quad \text{and} \quad m_t +P_2 \frac{\delta \mathcal{F} }{\delta m} =0\,,
\end{equation}
where 
\begin{equation}
P_1 = \partial_x ( m+\hat{k})+( m+\hat{k})\partial_x +  2\hat{k} \partial_x^{-1}\partial_y^2\, \quad P_2 = \partial_x-\partial_x^3\,.
\end{equation}
}
{\color{black}\begin{remark}
This Dirac reduction  procedure can be used to find  an asymptotic Hamiltonian structure obtained for the CH equation. The main difficulties are related to the CH structure, while the extension to the CH-KP equation is easily obtained from it through the addition of the term of order $\beta^2$ in the energy. More details on the 1D case are collected Appendix~B. 
\end{remark}}
{\color{black}\begin{remark}
The form of the Hamiltonian~\eqref{weaktrans} of the weakly transversal SGN model, and, in particular, the subsidiary role played by $\mu_2$,  suggests to consider the new coordinate set
\begin{equation}
\label{mcoords}
(\eta, \wit m, \mu_2)=(\eta, \eta\,\mu_1,\mu_2).
\end{equation} The Jacobian of this transformation is
\begin{equation}
J=\begin{pmatrix}
1&0&0\\
\mu_1&\eta&0\\
0&0&1
\end{pmatrix}\, , 
\end{equation}
and a straightforward computation shows that in the $(\eta,\wit m,\mu_2)$ coordinates, the Poisson tensor~\eqref{pxy}  acquires the form
\begin{equation}\label{npxy}
\wit{P}=-\begin{pmatrix}
\medskip
0&\partial_x \eta&\partial_y\\ \medskip
\eta\partial_x&\wit m\partial_x+\partial_x \wit m&\left(\dfrac{\wit m}{\eta}\right)\partial_y\\
\partial_y&\partial_y\left(\dfrac{\wit m}{\eta}\right)&0
\end{pmatrix}\, .
\end{equation}
If we consider the functional
\begin{equation}\label{Casemiro}
\mathcal{C}=\int 2\, \sqrt{\wit m}\, dx\, dy\ , 
\end{equation}
we get, calling $\delta$ the Lagrange differential, 
\begin{equation}
\label{PCase}
\wit{P}\delta{\mathcal{C}}=\wit{P}\begin{pmatrix}\medskip 0\\ \medskip \dfrac1{\sqrt{\wit m}}\\0
\end{pmatrix}
=-\begin{pmatrix}\medskip \left({\dfrac{\eta}{\sqrt{\wit m}}}\right)_x\\\medskip 0\\\left(\dfrac{\sqrt{\wit m}}{\eta}\right)_y
\end{pmatrix}
\end{equation}
since $\wit m \left(\dfrac1{\sqrt{\wit m}}\right)_x+(\sqrt{\wit m})_x=0$. Hence it makes sense to  consider the {\em Casimir manifold} defined by
\begin{equation}
\eta=\kappa\sqrt{\wit m}
\end{equation}
and follow the original steps in \cite{CH, CHH} to obtain the CH-KP hierarchy. In the core of the paper we preferred to follow the alternative path based on the coordinate change~\eqref{mu2ustar}.
\end{remark}}

%
%
%

 
\section{Some CH-KP solutions}
\label{Peaksol}

\begin{figure}[htbp]
    \centering
    \includegraphics[width=0.8\textwidth]{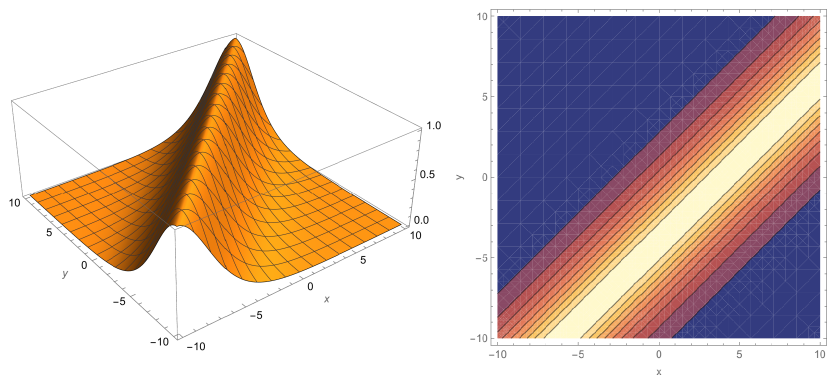}
    \caption{\color{black} An example of a smooth soliton for $\hat{k}=1$, $b=1$, and $\kappa=3/2$. Left panel: 3D rendition with axis rotation. Right panel: contour plot.}
    \label{fig:solitoniliscio}
\end{figure}

Travelling wave solutions of~\eqref{CH_KP} can be immediately obtained for the case of straight crests/troughs whereby the dependence on the $y$-coordinate enters linearly, 
\begin{equation}\label{tarveling}
U(x,y,t) = U(x -b y -c t)\, . 
\end{equation}
With respect to the one spatial dimension, the presence of the $U_{yy}$ term simply shifts the the term with the parameter~$\hat{k}$ from $2\hat{k}$ to $2\hat{k}(1+b^2)$ in the resulting ODE, so that after quadratures, with vanishing boundary conditions for $U$ and its derivatives as $x\to \infty$,  
this becomes just as in the 1D case, this equation defines a one-parameter solitary wave family with speed $c$ and amplitude
 $U_m=\sqrt{c^2-2 \hat{k}(1+b^2)}$, for $\hat{k}$ and $b$ fixed. 
 \begin{equation}\label{potential}
U_x^2 = U^2 \, \frac{c-2 \hat{k}(1+b^2)-U}{c-U}\, .
\end{equation}
An example of such solution in explicit form is 
$$
U(x,y,t) = {8 \over 3} \kappa \left( 
   1 - { 3 \sqrt{3} + 
       6 \sin(2 s) \over \big(1 + 2 \cos(2 s)\big) \big(2 \sqrt{3} \cos(2 s) - 
         \sqrt{3} \cos(4 s) + 2 \sin(2 s) + \sin(4 s)\big)} \,,
         \right)
$$
where 
$$s=
{1\over 3} \arctan(e^{(x - b y - c t)/2})\,, \qquad \kappa=\hat{k}+{b^2\over 2}\,,
$$
which represents a smooth soliton moving in the plane with speed $c=8 \kappa/3$ in the direction $\bm{\nu}=(1,-b)$ (Figure \ref{fig:solitoniliscio}). 
When $\hat{k}\to 0$ the amplitude of smooth solitary wave solutions of~\eqref{potential} tend to $c$ as the denominator in the equation cancels the numerator and the equation degenerates into 
\begin{equation}
U_x^2 = U^2 \,,
\end{equation}
while regularity is lost to developing a corner at the peak, 
\begin{equation}
U(x -b y -ct) = c e^{-|x-b y-ct|}\,.
\end{equation}
These were termed ``peakons" for the 1D case, and their proper mathematical interpretation is that of weak solutions of~\eqref{CH_KP}. 
 
 Owing to the fact that, in our scaling, the limit $\hat{k}\to 0$ essentially (after one integration with respect to $x$) reduces equation~~\eqref{CH_KP} to its one dimensional counterpart on the line,
\begin{equation}
\label{CH}
U_t - U_{xxt}+3 UU_x + (UU_{xxx}+2 U_xU_{xx})=0\, ,
\end{equation}
peakons play a similar role as for this 1D equation, i.e., they can be superimposed with time dependent amplitudes and speeds, the only difference being that their crests can now be tilted by an angle $\arctan(b)$ in the $(xy)$-plane.
Hence, the theory developed in~\cite{DCDS,CHL1,CHL2,HoldenRaynaud,CKL,Alina} and the ensuing peakon-based numerical algorithms for the integration of~\eqref{CH_KP} carry over to all initial conditions of ``tilted" kind
\begin{equation}\label{tiltic}
U(x,y,0) = U_0(x -b \, y)\, . 
\end{equation}
with $\kappa=\hat{k}(1+b^2/2)$. Specifically, as in 1D, peakons can be used as a non-orthogonal basis for integrable functions $U_0$ on the real line, with superposition coefficient whose time evolution yields a numerical approximation to the PDE solution:
\begin{equation}\label{N-soliton_sol}
U(x,y,t)=\sum_{i=1}^N p_i(t) \, e^{-|x-b \,y -q_i(t)|}\,,
\end{equation}
with the time evolution governed by the systems of ODEs 
\begin{equation}\label{sysNPeak}
\begin{split}
\dot{q}_{i}(t)&= \frac{1}{2}\sum_{j=1}^N p_j(t) e^{-|q_i(t)-q_j(t)|}-\kappa\, \\
\dot{p}_{i}(t) &=  \frac{1}{2}p_i(t)\sum_{j=1}^N p_j(t) \text{sgn} (q_i(t)-q_j(t)) e^{-|q_i(t)-q_j(t)|} \,.
\end{split}
\end{equation}
which is a Hamiltonian system of ODEs~\cite{CH,CHH,DCDS}.

Of course, the tilted initial conditions~\eqref{tiltic} and their evolution with parallel peakon crests are  straightforward generalizations of the 1D case, and the useful role played by peakons in constructing general solutions in that case may be lost in genuinely 2D setups, even under the rather unphysical 
limit $\hat{k}=0$. However, some peculiar behavior of certain solutions which are not trivially parallely tilted 1D setups is worth noticing. 

First, the 1D peakon-antipeakon collision~\cite{CH,CHH} can immediately be extended to a non-trivial 2D case~\eqref{CH_KP} by taking $\hat{k}=0$, 
\begin{equation}\label{papy}
U(x,y,t)=p(t)
\left(
e^{-|x- b\, y-q(t)|}-e^{-|x+b\, y+q(t)|} 
\right) 
\,, \qquad q(t)=\log(\text{sech}^2(t)) \,,\quad p(t)={c\over \tanh(t)} \,.
\end{equation}
Along the $x$-axis, this solution represents a collision of a peakon moving left to right with an antipeakon moving from right to left. Along $y=0$ the peakons are well separated as $t\to-\infty$ with their speeds and amplitude of magnitude $c>0$, and they annihilate each other, so that $U(x,0,0)=0$, at the collision time $t=0$, re-emerging at later times and exchanging their respective position.  At the moving point $x=0$, $y=-q(t)$ along the $y$-axis the crest and trough cross (while $U(0,y,t)=0$ is maintained for all $y$ and $t$). However, at generic positions in the plane the solution is different from zero, and the magnitude of crest and trough grows without bound as $t\to 0^-$. After the collision time the solution recovers finite magnitude as the peaks pass through each other along the $x$-axis. Note that if one were to use a different scaling making the coefficient of the $U_{yy}$ term independent 
of $\hat{k}$, then~\eqref{papy} would still be a solution for a particular (negative) choice of $\hat{k}$ that would cancel the contribution from that term with $2\hat{k} U_{xx}$. Similar considerations hold for the more general case of ``symmetric crossing" two-peakon transverse solutions, including the case of same-sign peak amplitudes.  
\begin{figure}[htbp]
    \centering
    \includegraphics[width=0.8\textwidth]{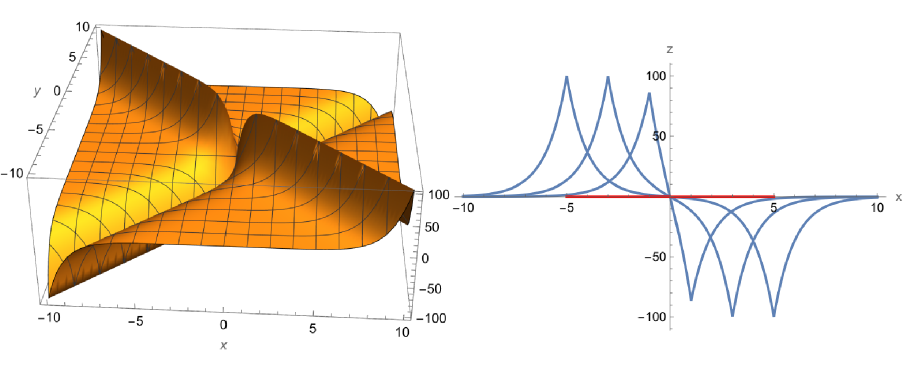}
    \caption{\color{black} Left panel: a three-dimensional view of the peakon profiles as $t\to0$. Right panel: cross-sections of the solution at different values of $y$: $y=0$ (red curve) and $y=1,3,5$ (blue curves). As $t\to0$, the peak amplitudes diverge for $y\neq0$.}

    \label{fig:solitoniespl}
\end{figure}
Next, more general 2D solutions based on peakons are also possible in the $\hat{k}=0$ case, as this limit reduces the role of the transverse $y$-coordinate to that of a mere parameter, since it is not involved in any derivative. Indeed, system~\eqref{sysNPeak} can be generalized to admit a $y$ dependence for the $q$'s and $p$'s. Just as in the 1D case, 
the equations of motion are given by the Hamiltonian 
\begin{equation}
H = \frac{1}{4}\sum_{i,j =1}^N p_i(y,t)p_j(y,t) e^{-|q_i(y,t)-q_j(y,t)|} \,,
\end{equation} 
\label{hamy}
so that 
\begin{equation}
\dot{q}_{i}= \frac{\partial H }{\partial p_i}\,, \qquad \dot{p}_{i} = -  \frac{\partial H }{\partial q_i}\,.
\label{sysqpy}
\end{equation}

The case $N=2$ again can serve as an illustration of these extended solutions. {\color{black}The ansatz solution in this case is 
\begin{equation}
U(x,y,t) = p_1(y,t)e^{-|x-q_1(y,t)|}+p_2(y,t)e^{-|x-q_2(y,t)|}\,,
\label{2pkgen}
\end{equation}
and following~\cite{CHH}, we assume
that as  $t\rightarrow \infty $, the solution asymptotically consists of two well-separated solitons with speeds $c_1, c_2$. }
These constants may depend on the parameter $y$. However, as in the 1D case, the dynamics change depending on whether $c_1>c_2>0$ or $c_2<0$, so that the $y$ dependence should respect one of these constraints. 
System~\eqref{sysqpy} has two constants of motion, 
\begin{equation}
P = p_1+p_2 = c_1+c_2\, ,
\end{equation}
the total momentum, and 
\begin{equation}\label{HAm-before_trans}
H = \frac{1}{2}(p_1^2+p_2^2)+p_1p_2 e^{-|q_1-q_2|} = \frac{1}{2}(c_1^2 +c_2^2)\, ,
\end{equation}
the Hamiltonian energy. 
 It is convenient to transform the coordinates from $p_i,q_i$ to 
\begin{equation}
P = p_1+p_2, \quad p= p_1-p_2, \quad Q = q_1+q_2, \quad q = q_1-q_2, 
\end{equation}
where we recall that these variables now include a parametric $y$-dependence.
The equations of motion are
\begin{equation}\label{system2peakons2d}
\begin{cases}
\dot{P} = 0\\
\dot{p}=\frac{1}{2}(P^2-p^2) \text{sgn}(q)e^{-|q|}\\
\dot{Q}= p(1+e^{-|q|})\\
\dot{q} =p (1-e^{-|q|})\\
\end{cases}
\end{equation}
generated by the Hamiltonian
\begin{equation}\label{H}
H = \frac{1}{2}(P^2+p^2)+\frac{1}{2}(P^2-p^2)e^{-|q|}\,.
\end{equation}
{\color{black} Notice that this Hamiltonian is twice the Hamiltonian energy \eqref{HAm-before_trans} of the system in the coordinates $p_i, q_i$, so from now on $H $ is a constant of  motion equal to $c_1^2+c_2^2$. 
The time evolution is obtained as in the 1D case~\cite{CH,CHH}, solutions are 
\begin{itemize}
\item[1)] \begin{equation}\label{c2 pos}
\begin{split}
q(y,t) &= \log \left(\frac{(c_1 c_2 e^{(c_1-c_2)t+d_0} + c_1^2)(c_1 c_2 e^{(c_1-c_2)t+d_0} + c_2^2) }{c_1 c_2 (c_1-c_2)^2 e^{(c_1-c_2)t+d_0}}\right)\,,\\
p(y,t) &= (c_1-c_2)\tanh \left(\frac{1}{2}((c_1-c_2)t+d_0)\right)\,,
\end{split}
\end{equation} 
when the parametric dependence of $c_1, c_2$ with respect to $y$ is such that $c_1>c_2>0 \quad \forall y \in \barr$
\item[2)] \begin{equation}
\begin{split}
q(y,t) &= \log \left|\frac{(c_1 c_2 e^{(c_1-c_2)t+d_0} -c_1^2)(c_1 c_2 e^{(c_1-c_2)t+d_0} - c_2^2) }{c_1 c_2 (c_1-c_2)^2 e^{(c_1-c_2)t+d_0}}\right|\,,\\
p(y,t)&=(c_1-c_2)\coth\left(\frac{1}{2}((c_1-c_2)t+d_0)\right)\,.
\end{split}
\end{equation}
when $c_1>0>c_2\,,\quad \forall y \in \barr$.
\end{itemize}
In both cases an initial shift $d_0$ (whose $y$-dependence can be general) enters the solution.
The original $(q_j,p_j)$-variables, $j=1,2$,  are
\begin{equation}
\begin{split}
p_1(t,y) &= \frac{P+p(t,y)}{2}\,,\quad p_2(t,y) = \frac{P-p(t,y)}{2}\,,\\ 
q_1(t,y)&= \frac{Q(t,y)+q(t,y)}{2}\,,\quad q_2(t,y)= \frac{Q(t,y)-q(t,y)}{2}\,.
\end{split}
\end{equation}}
{Representative examples to illustrate how different choices of the transverse dependence affect the geometry of the interaction can now be considered. 
Here we take  $c_1, c_2$ constants and show some examples with $d_0$ chosen as 
\begin{equation}
1)\,\,\,  d_0(y)\equiv \frac{1}{3}y\,,\quad 2)\,\, \, d_0(y)\equiv  \frac{1}{20}y^2\,.
\end{equation}
 {\color{black} With the first choice of $d_0$ correponds to cross shaped crests of peakons (Figures \ref{fig:solitoni1}, \ref{fig:solitoni2}). Unlike the peakon-antipeakon solution~\eqref{papy}, these never blow up. This is related to their construction, since here we extend via the parameter $y$ 1D solutions that  are always bounded. When $c_2$ is negative, there is a point of collision (Figure \ref{fig:solitoni1}). Conversely, when this parameter is positive, there is a point that minimizes the distance between the crests moving in time (Figure \ref{fig:solitoni2}). With the second choice of $d_0$, it is interesting to follow the evolution of the crests, since this  is not symmetric with respect to the direction of time. When time is negative the peakons collide at two points (case $c_2<0$, Figure \ref{fig:solitoni4}), or two points exist at which the distance between crests is minimal (case $c_2>0$, Figure \ref{fig:solitoni3}).  On the other hand, when time is positive, for both cases the distance between crests grows indefinitely.    }
\begin{figure}[htbp]
    \centering
    \includegraphics[width=\textwidth]{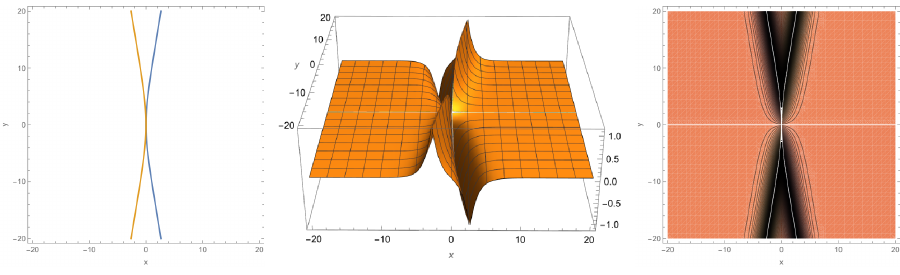}
    \caption{Two peakon solution~\eqref{2pkgen}. Left panel: crest location. Center panel: 3D rendition. Right panel:  level set contour plot. The initial chosen shift for this solution is $d_0 = \frac{1}{3}y$ while $c_1=1.2,\, c_2=-1.2$ are constants. {\color{black} The crest associated with the peakon $p_1e^{-|x-q_1|}$ is shown in blue in the left panel; as $y\to\infty$, this peakon propagates with asymptotic speed $c_1$ while the crest associated with $p_2e^{-|x-q_2|}$ is shown in yellow and has asymptotic speed $c_2$ as $y\to\infty$. When the two crests collide, the peakons exchange energy, giving rise to the observed phase shift. The variable $y$ plays a role analogous to time in this interaction.  
}}
    \label{fig:solitoni1}
\end{figure}
\begin{figure}[htbp]
    \centering
    \includegraphics[width=\textwidth]{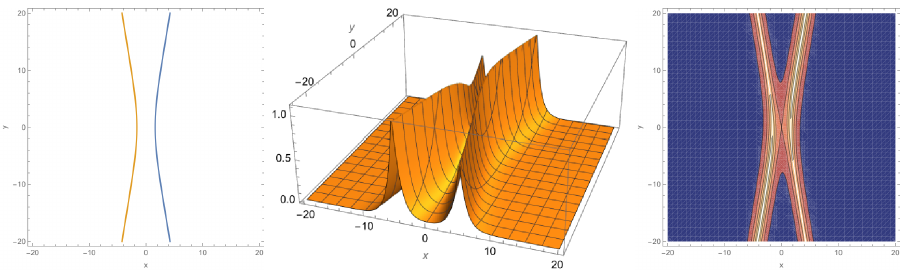}
    \caption{Same as Figure~\ref{fig:solitoni1}. Here the initial shift is $d_0 = \frac{1}{3}y$ while $c_1=1.2,\, c_2=0.8$ are constants. {\color{black} Left panel: the crest associated with the peakon $p_1e^{-|x-q_1|}$ is shown in blue; as $y\to\infty$, this peakon propagates with asymptotic speed $c_1$, while the crest associated with $p_2e^{-|x-q_2|}$ is shown in yellow and has asymptotic speed $c_2$ as $y\to\infty$. As the two crests approach each other, the peakons exchange energy, giving rise to the observed phase shift. Also in this interaction the variable $y$ plays a role analogous to time.}}
    \label{fig:solitoni2}
\end{figure}
\begin{figure}[htbp]
    \centering
    \includegraphics[width=\textwidth]{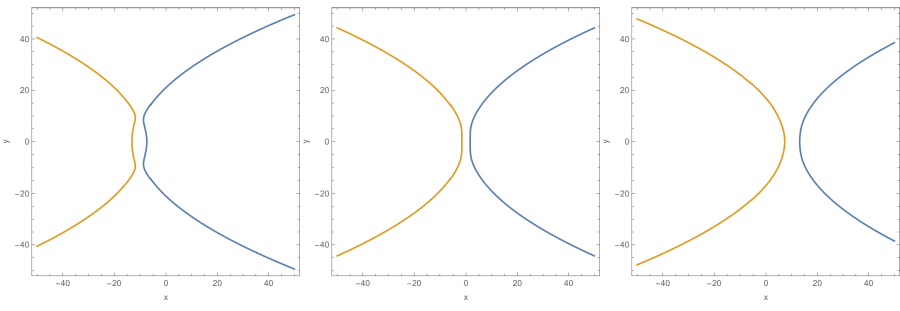}
    \caption{Crest locations of solution~\eqref{2pkgen}.  Shift is $d_0 = \frac{1}{20}y^2$,  $c_1 = 1.2, c_2 =0.8$. Left,  $t<0$, center, $t =0$, right, $t>0$. {\color{black} As in the previous figures the peakon $p_1 e^{-|x-q_1|}$ is shown in blue while $p_2 e^{-|x-q_2|}$ is shown in yellow. The parametric role of variable $y$ is similar to the other cases depicted above.}}
    \label{fig:solitoni3}
\end{figure}
\begin{figure}[htbp]
    \centering
    \includegraphics[width=\textwidth]{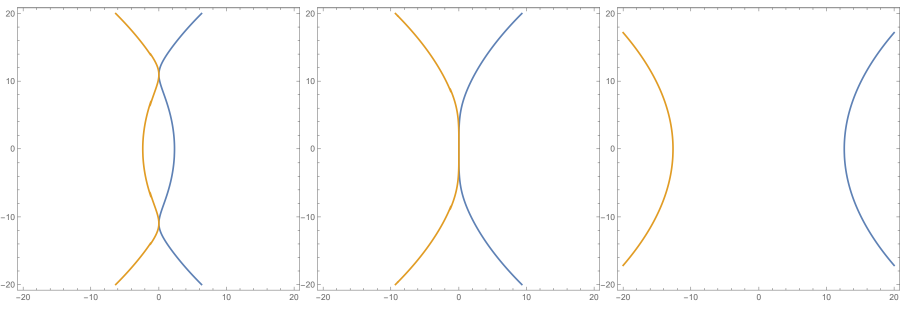}
    \caption{Same as Figure~\ref{fig:solitoni3}. The shift is $d_0 = \frac{1}{20}y^2$, while $c_1 = 1.2, c_2 =-1.2$. Left, $t<0$, center, $t =0$, right $t>0$. }
    \label{fig:solitoni4}
\end{figure}
\begin{figure}[htbp]
    \centering
    \includegraphics[width=0.8\textwidth]{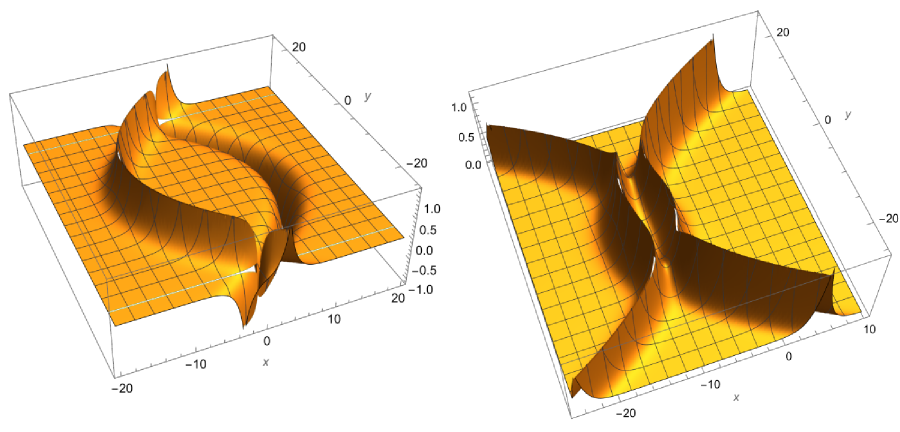}
    \caption{\color{black} 3D rendition of the solution with crest profiles shown in Figure~\ref{fig:solitoni3} (left) and~\ref{fig:solitoni4} (right) for $t<0$. }
    \label{fig:solitoni5}
\end{figure}

\section{Conclusions and perspectives}
In this work we focused on the Hamiltonian derivation of the CH-KP equation beginning from the Euler system for two-layer stratified fluids in three dimensional settings, next restricted by means of a double scaling limits to the classical free-surface water wave system {\color{black}(as done in two dimension in \cite{FS25})}. {\color{black}We have shown how the CH-KP  equation is the result of an asymptotic limit of long waves weakly dependent on one of the spatial coordinate. To this end, we introduced the horizontal momentum $\bs{\mu}$~\eqref{chieq} as the conjugate dependent coordinate to the layer thickness $\eta$ for the GN system, and we expressed it in terms of the horizontal velocity $u^*$~\eqref{mu2ustar} to find the unidirectional model~\eqref{CL11} with weak transversal $y$-dependence. 
The asymptotic expansion is augmented with a systematic Dirac approach so that the constrained Hamiltonian with the associated Poisson tensor are derived for the unidirectional model all the way from the parent Hamiltonian structure of the GN system.} The resulting CH-KP equation corresponds to the so called KP-II case and introduce a different, mildly more nonlinear balance with the dispersion of the latter. {\color{black} The CH-KP equation found from the asymptotic expansion can be transformed to the canonical \eqref{CH-KP} form through a suitable scaling of dependent and independent variables. In particular, we have considered one that is different than those in the current literature, see e.g. \cite{Guietal}. With this scaling, for the limit case $\hat{k}\rightarrow 0$ weak solutions extend directly to 2D the peakon case in one dimension.} We have provided a few examples of such solutions, but much remains to be done to interpret their role beyond that of mathematical interest, as is the case for the 1D where peakons can be used as a generalized bases to construct smooth solutions and follow their evolution by a particle based scheme \cite{CHL1, DCDS,CKL,Alina}. Stability issues of such and other solutions are important, but go beyond the scope of the present work and will be studied in the future. Notably, differences can be expected with the KP-I version of this equation, where some results have recently been obtained~\cite{Nilsson}. 
Future works on this subject will also extend the analysis to stratified fluids where KP-II has  also been derived, and in particular will concentrate on the role of the nonlocal operators  arising in this setting (\cite{BSL08,CFOPS25}).
}
\subsection*{Acknowledgments}
This study was carried out with support by the National Science Foundation under grants RTG DMS-0943851, CMG ARC-1025523, DMS-1009750, DMS-1517879, DMS-1910824, DMS-2308063, by the Office of Naval Research under grants N00014-18-1-2490, N00014-23-1-22478 and DURIP N00014-12-1-0749. 
The project has received fundings by the European Union’s Horizon 2020 research and innovation programme under the Marie Sk{\l}odowska-Curie grant no 778010 IPaDEGAN and by the PRIN 2022TEB52W-PE1 Project “The charm
of integrability: from nonlinear waves to random matrices”. We also gratefully acknowledge the financial
support of the GNFM Section of INdAM, and of the project MMNLP (Mathematical Methods in NonLinear Physics) of the INFN. RC thanks the Department of Mathematics and Applications of the University of Milano-Bicocca, ES thanks the Mathematics Department of UNC, and GF thanks SISSA-Trieste for their hospitality where part of this study was carried out.

\subsection*{Appendix A: The Dirac Reduction}\label{AppDirac}
\renewcommand{\theequation}{A\arabic{equation}}


{\color{black} 
As a side-product in his seminal paper on constrained Hamiltonian dynamics, Dirac~\cite{Dirac}  established that, 
 given an  $M$-dimensional  Poisson manifold $(\mathcal{M},  \{\cdot, \cdot\})$ and a submanifold $\mathcal{S}$  defined by the vanishing of $N$ independent constraints $\phi_a=0, a=1,\ldots, N$,  subjected to the  condition that the matrix $\mathsf{C}_{ab}=\{\phi_a,\phi_b\} $  be invertible the formula
 \begin{equation}
\label{DirFD}
\{g, f\}^D=\{g,f \}-\sum_{a,b=1}^N \{g, \phi_a\}(\mathsf{C}^{-1})_{ab} \{\phi_b, f\}\, , 
\end{equation} 
defines a new Poisson bracket on $\mathcal{M}$ which induces a corresponding Poisson structure (called Dirac-reduced Poisson structure) on $\mathcal{S}$.
 
An effective procedure to compute the Dirac reduced Poisson tensor goes as follows:
 \begin{enumerate}
\item Pick a set of coordinates adapted to the constraints, that is, consider the coordinate set $(y_{j_1}, \ldots, y_{_{M-N}}, \phi_1, \ldots, \phi_N)$, that is complement the constraints with independent coordinates $y_k, k=1,\ldots, M-N$.

\item Write the Poisson tensor $P$ corresponding to the original bracket $\{\cdot, \cdot\}$ in these new coordinates to obtain the new matrix representation
\begin{equation}
\label{PDtilde}
\wit{P}=\left( \begin{tabular}{c|c} $\Asf$&$\Bsf^T$
\\ \hline 
 $-\Bsf$&$\Csf$\smallskip\end{tabular}
\right)\, \,.
\end{equation}
\item Observe that the representation of the Dirac bracket~\eqref{DirFD} is given by
\begin{equation}
\label{DirRP}
\wit{P}^D=\left(
\begin{tabular}{c|c} $\mathsf{A}- \mathsf{B}\cdot\mathsf{C}^{-1}\cdot \mathsf{B}^T$&$0$
\\ \hline 
 $0$&$0$\smallskip\end{tabular}
%
\right)\, , 
\end{equation}
and in particular, that the constraint functions  $\phi_a$ are Casimir functions of~\eqref{DirFD}.
\item Finally, naturally obtain the Dirac reduced Poisson tensor in the free coordinates $(y_1, \ldots, y_{M-N})$ on $\mathcal{S}$ 
 by the $(M-N)\times(M-N)$  north-west block of \eqref{DirRP}, i.e., 
\begin{equation} \label{pdr}
\wit{P}^D_r=\Asf- \Bsf\cdot\Csf^{-1}\cdot \Bsf^T\, .
\end{equation}
\end{enumerate}
The advantage of such a procedure to obtain the reduced tensor is that it can be  transplanted to field-theoretic models (see, e.g., \cite{CFOPT23}, \cite{CFOPS25} and {\color{black} \cite{KacSolVal} for the case of Poisson vertex algebras}). 


\subsection*{Appendix B: Asymptotic reduction in the 1D case}}\label{1Dcase}
\renewcommand{\theequation}{B\arabic{equation}}
\setcounter{equation}{0}

The CH equation was first introduced in \cite{CH,CHH}through the reduction of the SGN system onto a Casimir manifold. Subsequently, it was also derived in \cite{CL09} {\color{black} and \cite{Johnson}} as a unidirectional model for the SGN system. Unidirectional models can be endowed with a Hamiltonian structure by means of a Dirac reduction and, as done for the 2D case in the main body  of the paper, a very similar procedure applies also in the 1D setting.
The starting point is the canonical Hamiltonian formulation of SGN system as obtained in \cite{FS25}, 
\begin{equation}
\label{AHeqm}
\left\{
\begin{array}{l}\smallskip
\dsl{\eta_t+(\eta\mu)_x+\frac{\eps^2 }{3}(\eta^3\mu_x)_{xx}}=0\,  \\
\dsl{\mu_t+\mu\mu_x+ g \rho\eta_x-\frac{\eps^2 }{2}(\eta^2{\mu_x}^2)_{x}}=0\, ,
\end{array}
\right.
\end{equation} 
where $\mu$ is related to the usual averaged velocity $\bar u$ through the asymptotic relation
\begin{equation}
\label {outomu}
\mu(\eta, \ou)=\rho\left( \ou-\dsl{\frac{\eps^2}{3\eta} (\eta^3\ou_x)_x}\right)\, .
\end{equation}
The energy  is 
\begin{equation}
\label{Haw}
\mathcal{H}[\eta,\mu]=\frac{1} 2\int_\RR \left((\eta\mu^2-\frac{\eps^2}{3}\eta^3{\mu_x}^2)+g(\eta-1)^2\right)\, d\, x\, ,
\end{equation}
while the canonical Poisson tensor is
\begin{equation}\label{Pred}
P{}
=-
\left(\begin{array}{cc}
0&\partial_x\\
\partial_x&0\end{array}
\right)\,  .
 \end{equation} 
\begin{rem}
Up to the change of variables \eqref{outomu}, the system \eqref{AHeqm} is the usual SGN model that is the starting point for the derivation of CH equation also in the seminal work \cite{CHH}.
\end{rem} 
Following \cite{CL09} we use the  velocity variable
\begin{equation}
u^*(x):= u\big(x, (1+\alpha \zeta(x))/\sqrt{{2}}-1)\big),
\end{equation}
where $u(x,z)$ is the horizontal component of the velocity vector.
By the usual Taylor expansion of the velocity potential, we can find an asymptotic relation between $u^*$ and $\mu$, given by 
\begin{equation}\label{u^*}
u^*= \mu + \frac{\epsilon^2}{4}\mu_{xx}+\alpha \epsilon^2\left(\zeta_x\mu_x+\frac{1}{2}\zeta \mu_{xx}\right).
\end{equation} 

This is exactly the same relation as the one obtained between $u^*$ and $\mu_1$ in the main body of the paper. The only difference is that, in the present 1D setting, there is no second component $\mu_2$. In section \ref{WTCHKP} we get rid of $\mu_2$ through the irrotationality property of the vector $\bs \mu$. Consequently, the only effect of $\mu_2$ was the appearance of the non locality in the Hamiltonian and in the equations of motion and did not otherwise affect the derivation of CH-KP equation. Therefore, in 1D it is possible to repeat the steps of Dirac reduction to find a 1D evolutive CH equation.

The SGN equations in the variables $(u^*, \zeta)$ up to order $\alpha \epsilon^2$ acquire the form
\begin{equation}
\label{GNuzeta}
\left\{
\begin{array}{l}\medskip
\zeta_t+u^*_x+\alpha (\zeta u^*)_x+\dfrac{\epsilon^2}{12}u^*_{xxx}+{\color{black}\alpha \dfrac{\epsilon^2}{4}(\zeta_x u^*_{xx}+\zeta u^*_{xx})}=0\\

u^*_t+\zeta_x+\alpha u^*u^*_x+\dfrac{\epsilon^2}{4}\zeta_{xxx}+{\color{black}\alpha \dfrac{\epsilon^2}{2}(2 u^*_xu^*_{xx}+2 \zeta_{x}\zeta_{xx}+\zeta \zeta_{xxx})}=0\, .
\end{array}
\right.
\end{equation}
and its Hamiltonian structure can be read from the formulas \eqref{nHtwas}, setting $\beta =0$, and \eqref{npoit} restricted to the principal minor of order 2 in the upper-left corner.

%
%
We obtain a unidirectional model by imposing the
(asymptotic) relation
 \begin{equation}\label{2dcostr}
 \zeta -( u^* +\frac{\alpha}{4}{u^*}^2 -\frac{\epsilon^2}{12}u^*_{xx}-\alpha \epsilon^2 \left(\frac{17}{48}{u^*_{x}}^2+\frac{5}{24}u^*u^*_{xx}\right)=0\, .
 \end{equation}
From now on, the computations to find the reduced Poisson tensor are identical to the ones of Section \ref{WTCHKP}, we consider the set of coordinates $( u^*,\Phi)$, where $\Phi$ is defined by the LHS of \eqref{2dcostr} and then we find $P_{red}$ applying the Dirac procedure to be exactly \eqref{redtensor}. 

The Hamiltonian  restricted to the constraint \eqref{2dcostr} is 
\begin{equation}
\mathcal{H}= \int \Big({u^*}^2 + \alpha \frac{3}{4}{u^*}^3 +\frac{\epsilon^2}{6} {u^*_x}^2 +\alpha \epsilon^2 \frac{3}{16}{u^*_x}^2 u^*\Big) dx\, .
\end{equation}
The reduced Poisson structure gives the 1D local evolution equation, 
\begin{equation}\label{CL1}
u^*_t +u^*_x +\frac{3}{2} \alpha u^*u^*_x+\frac{\epsilon^2}{6}u^*_{xxx}+\alpha \epsilon^2\left(\frac{35}{24} u^*_x u^*_{xx}+\frac{5}{12}u^* u^*_{xxx}\right)=0
\end{equation}
which coincides with the one obtained in ~\cite{CL09} by a direct inspection of \eqref{GNuzeta}.

To obtain the CH equation,  we observe that the Hamiltonian evolution equation \eqref{CL1} is asymptotically equivalent to 
\begin{equation}\label{CL3}
u^*_t -\frac{5}{12}\epsilon^2 u^*_{xxt}+u^*_x +\frac{3}{2} \alpha u^*u^*_x-\frac{\epsilon^2}{4}u^*_{xxx}-\alpha \epsilon^2 \frac{5}{24}(2 u^*_x u^*_{xx}+u^* u^*_{xxx})=0\,.
\end{equation}
To arrive at the canonical form of the CH equation one  performs the scaling
\begin{equation}
u^*(x,t)= a U(b(x-vt),ct)
\end{equation}
with $a,b,c$ that are as in \eqref{abc}, the new variable $U$ satisfies 
\begin{equation}
U_t- U_{xxt} +2\hat{k}U_x+3  UU_x = UU_{xxx}+2U_x U_{xx}\, .
\end{equation}

The purpose of this appendix is to show that the Dirac reduction carried out in the main body of the paper is a natural extension of the corresponding one-dimensional construction. For readability, however, we preferred to present all computations directly in the two-dimensional setting in Section \ref{WTCHKP}. This asymptotic reduction in 1D is a different way to endow the CH equation with a Hamiltonian structure with respect to \cite{CH, CHH}, even if it is clear that the resulting equation differs of a scaling of the coefficients. More importantly, the two-dimensional framework developed in the main text naturally leads to the CH--KP equation, which is the main result of the paper.

\end{document}